\documentclass[10pt,twocolumn,letterpaper]{article}

\usepackage{wacv}
\usepackage{times}
\usepackage{epsfig}
\usepackage{graphicx}
\usepackage{amsmath}
\usepackage{amssymb}
\usepackage{booktabs}
\usepackage{algorithm}
\usepackage{algorithmic}

\usepackage[accsupp]{axessibility}  

\def\wacvPaperID{1289} 

\wacvapplicationstrack 

\wacvfinalcopy 

\ifwacvfinal
\usepackage[breaklinks=true,bookmarks=false]{hyperref}
\else
\usepackage[pagebackref=true,breaklinks=true,colorlinks,bookmarks=false]{hyperref}
\fi

\begin{document}

\title{Performer: A Novel PPG-to-ECG Reconstruction Transformer for a Digital Biomarker of Cardiovascular Disease Detection}

\author{Ella Lan\\
%
}

\maketitle
\thispagestyle{empty}

\begin{abstract}
   Electrocardiography (ECG), an electrical measurement which captures cardiac activities, is the gold standard for diagnosing cardiovascular disease (CVD). However, ECG is infeasible for continuous cardiac monitoring due to its requirement for user participation. By contrast, photoplethysmography (PPG) provides easy-to-collect data, but its limited accuracy constrains its clinical usage. To combine the advantages of both signals, recent studies incorporate various deep learning techniques for the reconstruction of PPG signals to ECG; however, the lack of contextual information as well as the limited abilities to denoise biomedical signals ultimately constrain model performance. In this research, we propose Performer, a novel Transformer-based architecture that reconstructs ECG from PPG and combines the PPG and reconstructed ECG as multiple modalities for CVD detection. This method is the first time that Transformer sequence-to-sequence translation has been performed on biomedical waveform reconstruction, combining the advantages of both PPG and ECG. We also create Shifted Patch-based Attention (SPA), an effective method to encode/decode the biomedical waveforms. Through fetching the various sequence lengths and capturing cross-patch connections, SPA maximizes the signal processing for both local features and global contextual representations. The proposed architecture generates a state-of-the-art performance of 0.29 RMSE for the reconstruction of PPG to ECG on the BIDMC database, surpassing prior studies. We also evaluated this model on the MIMIC-III dataset, achieving a 95.9\% accuracy in CVD detection, and on the PPG-BP dataset, achieving 75.9\% accuracy in related CVD diabetes detection, indicating its generalizability. As a proof of concept, an earring wearable named PEARL (prototype), was designed to scale up the point-of-care (POC) healthcare system.
\end{abstract}

\section{Introduction}

Cardiovascular diseases (CVDs) have become the leading cause of death. According to the World Health Organization, roughly 17.9 million people died from CVDs in 2018, with three-quarters of these deaths occurring in lower-income communities; a primary reason is inaccessible on-demand healthcare monitoring infrastructure. To tackle this issue, in the past several years, many consumer-grade wearables have been developed to provide home-based solutions for daily cardiac monitoring through the measurement of ECG and PPG.

Electrocardiography (ECG) is a widely used gold standard for cardiovascular diagnostic procedures ~\cite{somani2021deep}. It contains five peaks, P, Q, R, S, and T, with each component representing a different electrical recording of the heart’s activity. ECG can be used to evaluate different metrics of the heart’s functionality ranging from heart rhythm to the strength of each impulse, providing critical information for diagnosing numerous diseases such as enlargement of the heart, congenital heart defects, and heart inflammation. Thus, ECG monitoring is proven to be beneficial for early detection of CVDs, which is essential for increasing the chances of survival in patients, especially those with a high risk or in the aging population. However, the most conventional ECG-based wearables for daily tracking may restrict users’ activities, making it infeasible to use ECG for continuous cardiac monitoring. For instance, the Apple Watch, a popular wrist-based ECG monitoring solution, requires users to keep their hands on the watch, only measuring the signals for up to 30 seconds. ECG-based solutions require the “closing of the circuit”, which reduces the convenience of daily tracking. More fundamentally, cardiac monitoring with a limited duration leads to the possibility of missing key asymptomatic and irregular signals. 

On the other hand, photoplethysmography (PPG) is an optically obtained signal that can be used to detect blood volume changes in the microvascular bed of tissues. This method can be used to extract key information such as blood oxygen saturation, heart rate, blood pressure, cardiac output, respiration, arterial aging, endothelial function, microvascular blood flow, and autonomic function ~\cite{elgendi2019use}. Compared to ECG, PPG is convenient to set up, cost-effective, and less invasive. It is also more user-friendly for long-term continuous cardiac monitoring as there is no requirement for constant user participation. However, the quality of the measured PPG signals fluctuates notably because of the influence of irrelevant outside factors, such as the individual’s skin color and motion artifacts. Recently, studies have further analyzed the applications of PPG, suggesting that PPG is an effective non-invasive digital biomarker for CVD-related diseases such as diabetes. Avram \etal~\cite{avram2020digital} conducted a large-scale study analyzing the functionality of PPG using CNN; it was concluded that (1) PPG contains clinically valuable information and (2) despite the noise in data, its usage can be further extended through the application of deep learning.

PPG and ECG are physiologically related as they embody the same cardiac process in two different signal-sensing domains ~\cite{gil2010photoplethysmography}. The peripheral blood volume change recorded by PPG is heavily influenced by the contraction and relaxation of the heart muscles, which is controlled by the cardiac electrical signals triggered by the sinoatrial node. The intrinsic correlation between PPG and ECG fundamentally inspires this research, leading to the creation of a novel digital biomarker, PPG with its reconstructed ECG, that combines the benefits of both signals: the convenience of PPG and the accuracy of ECG. This enables an effective and continuous cardiac monitoring solution that would reconstruct the ECG from the PPG and connect the rich information of both signals for CVD detection. To perform the waveform reconstruction and CVD detection, this paper proposes a Transformer-based architecture, the \textbf{P}PG-to-\textbf{E}CG \textbf{R}econstruction Trans\textbf{former} (Performer), enabling the creation of the described digital biomarker.

\begin{figure}
\begin{center}
\includegraphics[width=0.99\columnwidth]{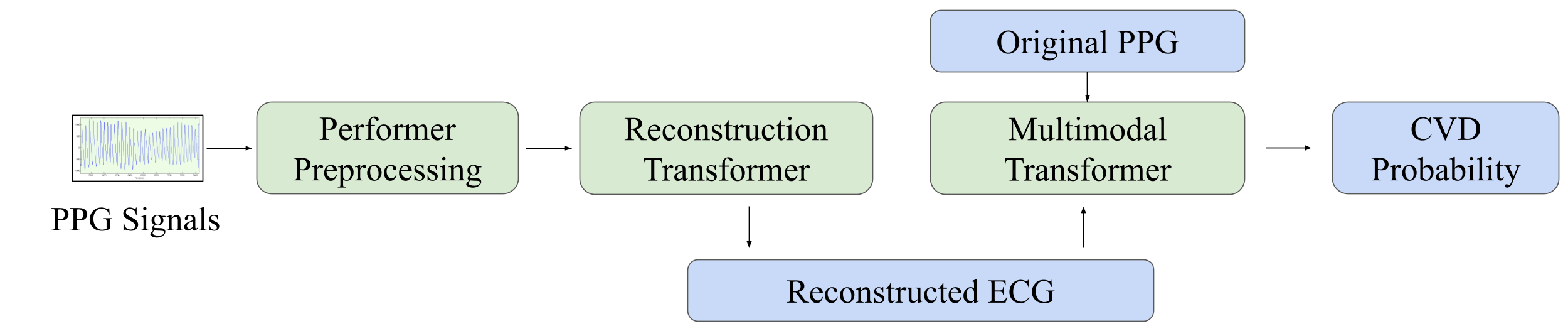}
\end{center}
   \caption{Overall Architecture of Performer.}
\label{figure1}
\end{figure}

In this work, we use Transformer as the backbone of the Performer architecture. In addition, we design the Shifted Patch-based Attention (SPA) mechanism to optimize the encoding/decoding process for the biomedical waveforms: (1) a set of various sized patches are used to fetch the different sequence lengths into Performer to capture the signal features, and (2) a shifted patch mechanism is devised to enable the attention among the patches to cover cross-patch connections. In summary, the main contributions of this work are:

\begin{itemize}
    \item We introduce a Transformer-based architecture (Performer) to successfully reconstruct the ECG signals using only the PPG input; this approach enables continuous cardiac monitoring through combining the advantages of both signals. The architecture also offers a solution to use the PPG and reconstructed ECG as multimodalities for CVD detection. To the best of our knowledge, this was the first time that Transformer has been used for biomedical waveform reconstruction.
    \item We propose the SPA mechanism that integrates various patch sizes as multiple hierarchical stages to capture specific aspects of PPG/ECG, hence providing an effective and universal approach to process biomedical waveforms.
    \item We create a new digital biomarker, PPG with its reconstructed ECG, for CVD detection.
    \item We prototype an earring wearable as a proof of the concept for effective continuous cardiac monitoring in daily life. 
\end{itemize}

\begin{figure}[t]
\begin{center}
\includegraphics[width=0.9\columnwidth]{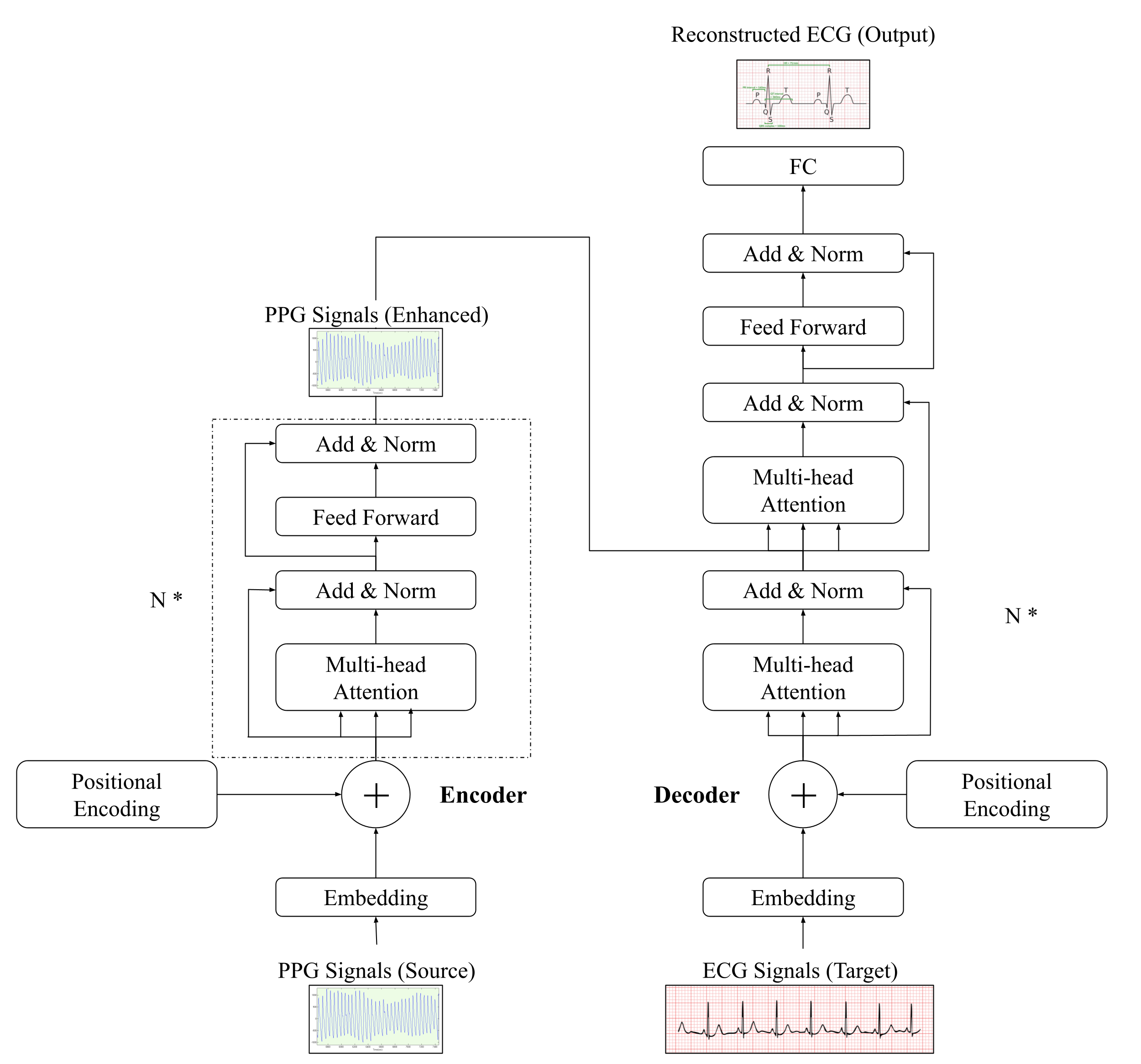}
\end{center}
   \caption{PPG-to-ECG Reconstruction Encoder / Decoder.}
\label{figure2}
\end{figure}

\section{Related Work}

\subsection{PPG-to-ECG Reconstruction}

Research on the PPG-based ECG inference is in its infancy; few prior studies have been dedicated to this topic. The preliminary studies in this field used feature-based machine learning. Banerjee \etal~\cite{banerjee2014photoecg} created a computational parameter model that extracted features from PPG, such as pulse transit time (PPT), to predict ECG parameters. Zhu \etal~\cite{zhu2021learning} used a mathematical approach, creating a discrete cosine model to study correlations between PPG and ECG. However, gaps arise in analyzing limited features: the lack of non-linear information representation reduces the models’ accuracy of ECG reconstruction and limits applicability to real medical cases.

With the recent rise of deep learning, this problem was initially addressed through models based on CNN; signal processing was performed by localizing key features and reducing noise for both PPG and ECG. Avram \etal~\cite{avram2020digital} utilized smartphone-based PPG signals and CNN to achieve an area under curve (AUC) of 0.75 with a confidence level of 95\% for diabetes prediction. However, the lack of contextual information and long-range dependencies by locality sensitivity were limits acknowledged after experimentation. Chiu \etal~\cite{chiu2020reconstructing} used recurrent neural networks (RNN) with an encoder/decoder architecture for sequence-to-sequence learning, but the model’s sensitivity to signal noise and gradient-descent approach limits its performance on tasks involving long-term memory. Sarkar \etal~\cite{sarkar2021cardiogan} applied generative adversarial networks (GAN) with a generator/discriminator architecture to generate artificial data as data augmentation for the synthesis of ECG from PPG. However, the challenge of overcoming unstable training resulted in occasional but critical random oscillations.

\subsection{Transformers and Self-Attention Mechanism}

Although Transformer originated in the world of natural language processing (NLP) ~\cite{vaswani2017attention}, it has also become prevalent in the field of computer vision (CV) and has surpassed many CNN-based models for tasks such as image classification and segmentation. Much of Transformer’s success comes from its self-attention mechanism ~\cite{bahdanau2014neural}, which not only simplifies the architectural complexity by removing convolutions entirely, but also allows models to capture the global contextual information for both short-range and long-range relationships. Even after the development of Vision Transformer (ViT) ~\cite{dosovitskiy2020image}, advancements in the field of Transformers have been continuously made. Carion \etal~\cite{carion2020end} continued to advance Transformers in the CV world through their creation of the Detection Transformer (DETR). DETR performed on par with the well-established Faster R-CNN in terms of run-time and accuracy and outperformed the SOTA Faster R-CNN in generalizability. Liu \etal~\cite{liu2021swin} created the Shifted Window Transformer (Swin Transformer) as an improved vision Transformer, creating a hierarchical structure that implements a shifted window approach. 

As Transformer gains popularity, the application of its self-attention mechanism has been extended to many fields. The specific task in this work is to create a Transformer encoder/decoder architecture that uses sequence-to-sequence processing for biomedical waveform reconstruction. We focus on effective tokenization for biomedical waveforms to maximize the local features and global contextual representations, producing consistent training results universally.

\begin{figure}[t]
\begin{center}
\includegraphics[width=0.9\columnwidth]{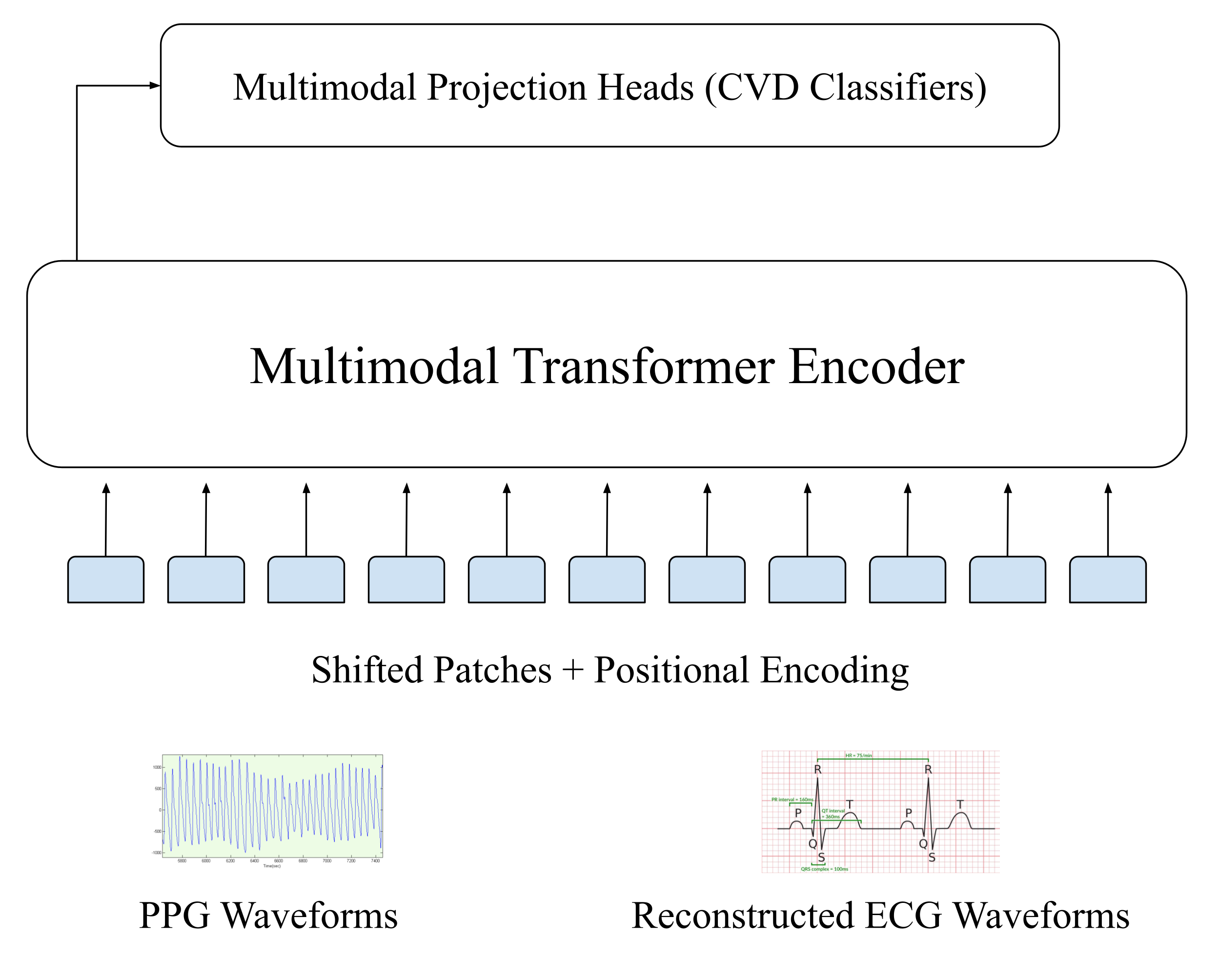}
\end{center}
   \caption{Multimodal Architecture of Performer.}
\label{figure3}
\end{figure}

\section{Method}

\subsection{Objectives and Design}

In addition to the incorporation of Transformer, current clinical methods for biomedical waveform analysis and disease diagnosis inspire us to design various aspects of the proposed deep learning architecture.

\begin{figure*}
\begin{center}
\includegraphics[width=0.99\textwidth]{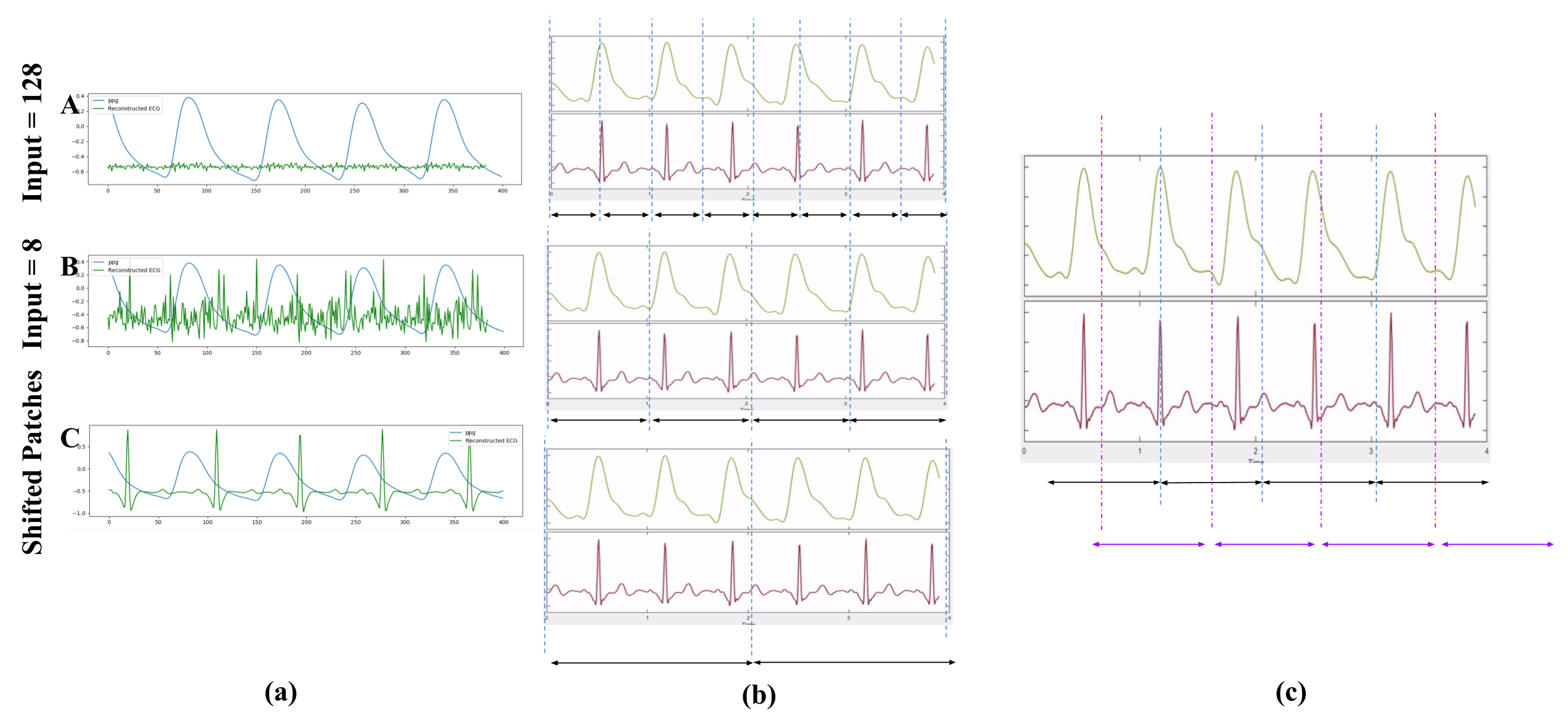}
\end{center}
   \caption{Shifted Patch-based Attention Design. (a) Comparison and Experiments of SPA. (b) Examples of the Different Patch-size Tokenizations. (c) Example of the Shifting Mechanism.}
\label{figure4}
\end{figure*}

For accurate analysis of medical waveforms, especially for PPG (given its susceptibility to motion, light, skin type, etc.), signal denoising becomes critical. Accordingly, Transformer’s self-attention module assigns weights of importance to sequences proven to contain more valuable contextual information, using the defined weights to identify whether each section of the sequence should be focused on or filtered out. This approach enables the Performer architecture to automatically denoise the PPG and allocate attention to key features, such as the PQRST waves in the ECG complex. 

\begin{figure}[t]
\begin{center}
\includegraphics[width=0.9\columnwidth]{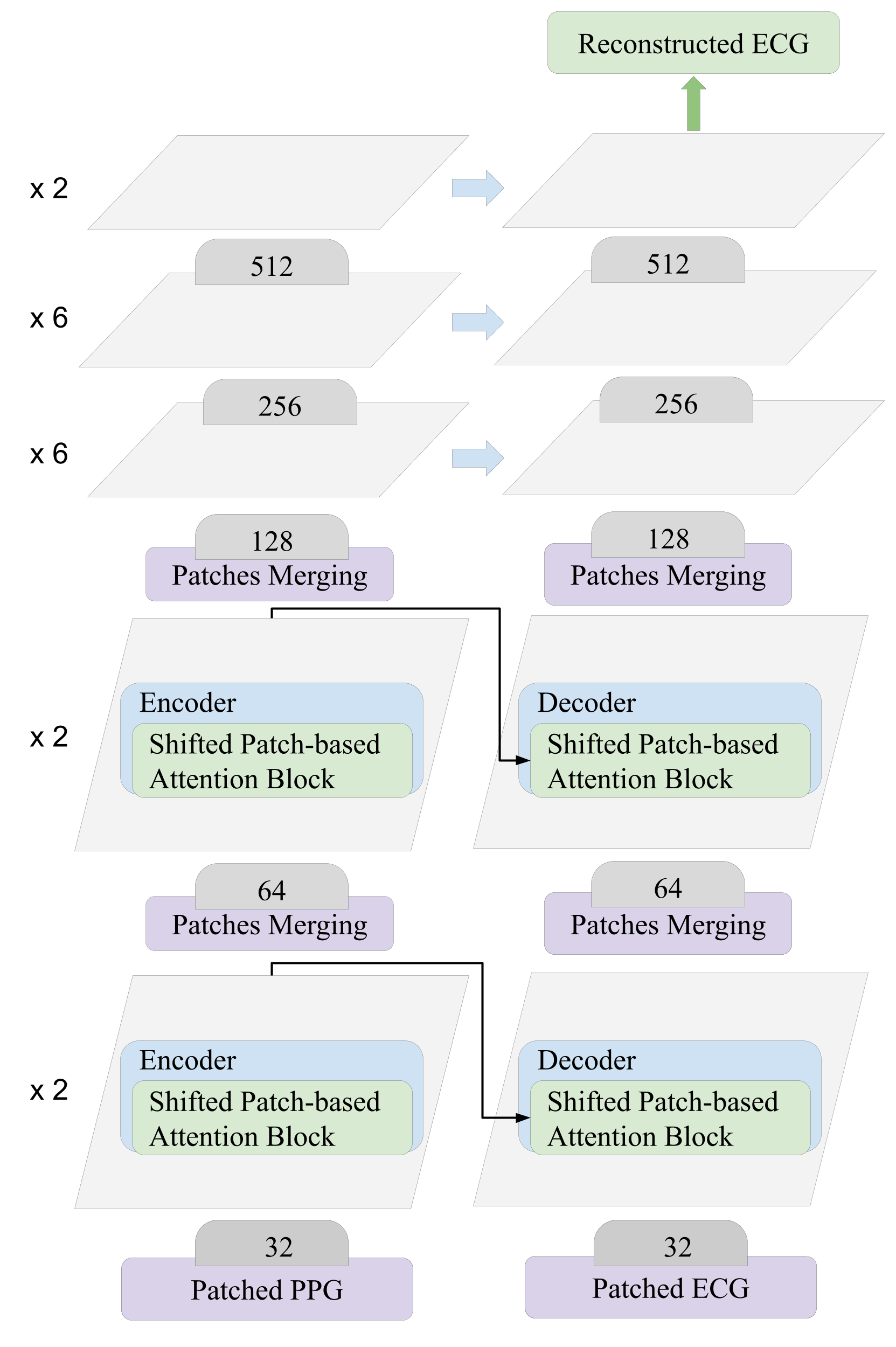}
\end{center}
   \caption{Shifted Patch-based Attention Framework.}
\label{figure8}
\end{figure}

To measure cardiac activities, clinical analyses of ECG/PPG signals evaluate multiple factors such as amplitude, time interval, direction, scale, moving connection, angle, and shape. Hence, positional embeddings and multi-head attention are incorporated into Performer to accurately capture the key signal features and provide the contextual relationships between each medical waveform. In this study, we also implement sequence-to-sequence training to analyze short/long-range signals and create a multimodal training procedure to incorporate the PPG and reconstructed ECG for CVD detection. Performer uses its hierarchical design ~\cite{lin2017feature} to handle non-linear complex information representation.

\subsection{Proposed Architecture}

Figure~\ref{figure1} shows the three key parts of Performer: the initial data preprocessing of the PPG and ECG waveforms, the reconstruction Transformer, and the multimodal Transformer. First, the raw PPG signals are fed into the waveform data preprocessing module for denoising, standardization, and normalization. The reconstruction Transformer module then produces the reconstructed ECG signals and combines them with the original PPG signals for the multimodal Transformer module, leading to the model’s CVD classification. 

\subsubsection{Reconstruction Transformer}

As illustrated in Figure~\ref{figure2}, a Transformer encoder/decoder architecture with its attention mechanism is utilized to reconstruct ECG from PPG. For training, both PPG signals (source) and paired ECG signals (target) are fed into the reconstruction Transformer as the 1D time series input, generating the reconstructed ECG signals as the output. Incorporating the positional embeddings, Performer’s multi-head attention mechanism for sequence-to-sequence training computes the relationships of each sequence to all others in the allotted window. Based on the contextual relationships, the model assigns weights to different subsections of each waveform, focusing on key signals, while also performing an indirect denoising process. The attention calculation is described as

\[ Attention(Q, K, V) = Softmax \left( \frac{Q K^{T}}{\sqrt{d_{k}}} \right) V \]

where Q, K, V denote Query, Key, and Value, as defined by Transformer’s self-attention.

Since this proposed method is an end-to-end training without complicated handcrafted hyper-parameters to calculate the reconstruction, a simple loss function is chosen as follows: 

\[ Loss = \sum_{i=1}^{N}{|y_i - \acute{y_i}|} \]

where \({y_i}\) denotes the ground truth value of ECG, \(\acute{y_i}\) represents the reconstructed ECG value predicted at position i, and N represents the total number of sampling values for each original PPG/ECG waveform segmentation.

\subsubsection{Multimodal Transformer}

Given the significance of PPG and ECG waveforms for cardiovascular monitoring, Performer utilizes its multimodality capability to build a multimodal Transformer (Figure~\ref{figure3}) which takes inputs from both the PPG and reconstructed ECG signals for CVD detection. Each modality representation is added into a feature vector and fed into the SPA-based multimodal Transformer encoder. Through the multimodal Transformer, Performer analyzes the connections between the PPG and ECG signals in addition to the relationships among sequences within the same waveform. The projection heads are added to the last layer of the encoder as the CVD classifier. This design is flexible to adopt other types of vital signs, such as age and weight, as text-based encoders, to continue enhancing model performance.

The model hyperparameters and layers in both the reconstruction Transformer and multimodal Transformer will be described in the following as part of the SPA framework.

\subsection{Shifted Patch-Based Attention}

There is minimal existing research applying Transformers on waveforms; effective tokenization is at the frontier of new research. Many recurring errors arose initially when inputting waveforms of a fixed length into the reconstruction Transformer model. Compared with words or texts (the usual Transformer inputs), biomedical waveforms result in significant differences, due to the larger variation in the scale of waveforms and the length of the data input to be tokenized.

\begin{algorithm}[tb]
\caption{Shifted-Patched based Attention Algorithm}
\label{alg:algorithm1}
\textbf{Input}: Biomedical Waveforms\\
\textbf{Parameter}: Patch Size, Neural Network Layer Config\\
\textbf{Output}: Hierarchical Tokenized Waveform Representation
\begin{algorithmic}[1] 
\STATE Initial patch partition.
\WHILE{current stage within the layer config}
\IF {current stage input != initial patched embedding}
\STATE Perform patches merging.
\ENDIF
\STATE Perform source features extraction via Encoder.
\IF {reconstruction flag == true}
\STATE Perform target reconstruction via Decoder.
\ENDIF
\STATE Advance to next stage.
\ENDWHILE
\STATE \textbf{return} SPA-based waveforms
\end{algorithmic}
\end{algorithm}

\subsubsection{Hierarchical Architecture}

To capture the key patterns and relationships, the tokenization of waveforms is critical for the embedding. A universal solution is required to optimize the performance. In Figure~\ref{figure4}(a), the blue waveform represents the PPG input, and the green waveform represents the ECG signal prediction of the three trials. In panel (A), a fixed-length input of 128 points (a relatively long sequence length) generated narrow range curves. In panel (B), a fixed-length input of 32 points (a relatively shorter sequence length) led to stochastic predictions. The initial prediction outcomes indicated that different lengths of the same waveform led to different observations by the model, which made the vanilla Transformer model inadequate to analyze a fixed patch size of biomedical waveforms. To address the issue of the standard Transformer, SPA was created and – as seen in panel (C) – led to a substantial increase in ECG reconstruction accuracy, by fetching various sequence lengths and covering cross-patch connections.

\subsubsection{Various Patch Size Merging}

SPA’s architecture consists of a set of patch-based algorithms implemented at multiple hierarchical stages. The waveforms are first pre-processed and converted into a 1D vector containing the raw values of each sequence, which is then split in various lengths of patches in the different stages. At the initial patch split, each patch has the size of 32 value points as the raw-value features; these patches are fed into the Transformer encoder/decoder as tokens. Then, in each stage, a patch-merge operation is performed to reduce the number of the tokens by half while maintaining the dimension for each token. It produces the hierarchical stages illustrated in Figure~\ref{figure8}, covering the different wave lengths of 32 points, 64 points, 128 points, 256 points, and 512 points, respectively. Each stage contains pairs of encoders and pairs of decoders, with the model layer numbers as \{2, 2, 6, 6, 2\}. Figure~\ref{figure4}(b) shows an example of the same waveform being split into different patch sizes according to the stage level of SPA, in which the green waveform represents PPG and the red waveform represents ECG.

\begin{table*}
  \begin{center}
    {\small{
\begin{tabular}{c|c|c|c|c|c|c}
\toprule
Model & DCT & XDJDL & CardioGAN & Encoder/Decoder & Vanilla Transformer & Performer \\
\midrule
RMSE & 0.67 & 0.39 & 0.36 & 0.34 & 0.51 & \textbf{0.29} \\
\bottomrule
\end{tabular}
}}
\end{center}
\caption{RMSE Comparison with SOTA Results.}
\label{table1}
\end{table*}

\subsubsection{Shifting Patch Mechanism}

In each stage, there are pairs of encoders and pairs of decoders as attention blocks: on two successive attention blocks, one block is used to analyze the original sequence corresponding to the patch size, while the other block represents the sequence resulting from the shifted mechanism of SPA. As can be seen in Figure~\ref{figure4}(c), the “shifted” operation shifts the start point of each patch by half of the patch size. Since it is essential to capture the cross-patch connections between tokens, a shifted patch mechanism allows the attention to focus on key segments between patches. Each SPA block and its consecutive SPA block are computed as 

\[ \acute{z}^l = PA(LN(z^{l-1})) + z^{l-1} \]
\[ z^l = MLP(LN(\acute{z}^l)) + \acute{z}^l \]
\[ \acute{z}^{l+1} = SPA(LN(z^l)) + z^l \]
\[ z^{l+1} = MLP(LN(\acute{z}^{l+1})) + \acute{z}^{l+1} \]

where \(\acute{z}^l\) denotes the features output from the original patch and \(z^l\) denotes the features output from the shifted patch at Stage l. PA stands for the initial patch attention, SPA stands for the shifted patch attention, LN stands for layer normalization, and MLP stands for multilayer perceptron.

The full SPA approach with the detailed implementation is illustrated in Algorithm~\ref{alg:algorithm1}.

\begin{figure}[t]
\begin{center}
\includegraphics[width=0.9\columnwidth]{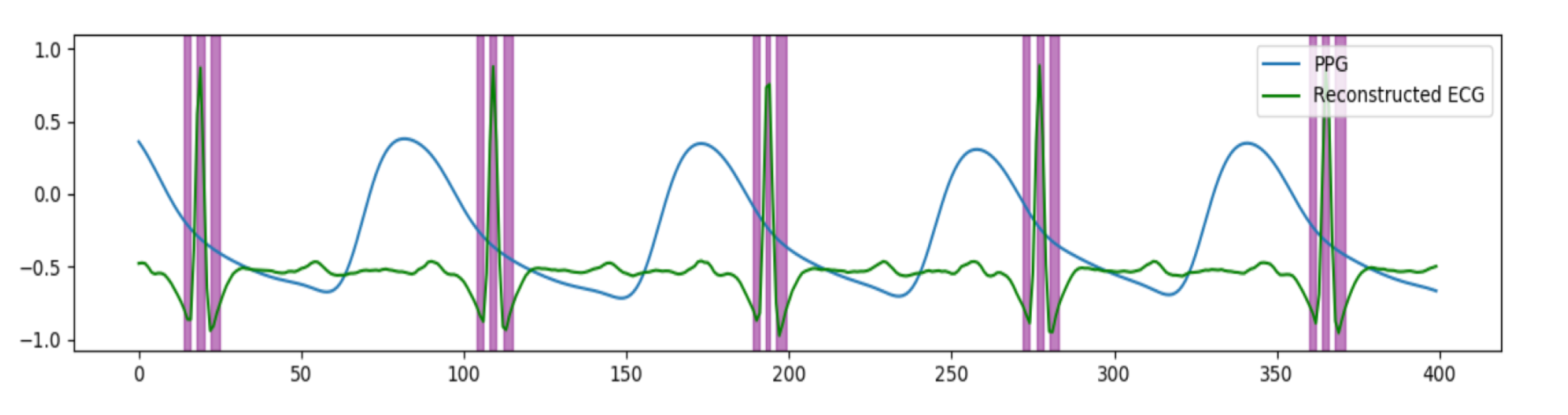}
\end{center}
   \caption{Attention Maps through SPA.}
\label{figure5}
\end{figure}

\section{Experiments}

\subsection{Datasets}

There are few publicly available datasets that contain both paired PPG/ECG signals and their corresponding CVD diagnosis. To overcome this issue, we use two categories of datasets in this research.

\subsubsection{PPG Paired with ECG}

The first type of dataset contained pairs of PPG signals and Lead-II ECG signals but without CVD labels. The UQVSD dataset had roughly 3000+ minutes ~\cite{liu2012university}, DaLiA had roughly 1800+ minutes ~\cite{reiss2019deep}, and BIDMC had roughly 400+ minutes ~\cite{goldberger2000physiobank}. These datasets generated a total of 5200+ minutes of PPG signals paired with ECG signals, which were used to train, validate, and test the reconstruction Transformer. The main reason for selecting these datasets was their inclusion of Lead-II ECG. Among all other leads, we focused on reconstructing Lead-II ECG due to its convenient location and vital cardiovascular information. Lead-II is often used because it provides the best view of the heart conduction system activity (P and R waves). Lead-II lies closest to the cardiac axis and is right-arm negative, left-leg positive, meaning it shows the impulse traveling from the right atria to the left ventricle.

\subsubsection{PPG Paired with Corresponding CVD}

After the optimization of Performer’s reconstruction module, the multimodal Transformer was refined through the usage of datasets that contained PPG signals paired with the corresponding CVD diagnosis. The MIMIC-III dataset (40,000 patients) ~\cite{johnson2016mimic} was chosen; it consisted of PPG signals paired with one of the four following CVDs: coronary artery disease (CAD), congestive heart failure (CHF), myocardial infarction (MI), and hypotension (HOTN). Its distributions of class labels are 7:7:7:12. The PPG signals were first inputted into the reconstruction Transformer to generate the reconstructed ECG. Provided that the PPG signals had their corresponding CVD labels and the reconstructed ECG was effectively generated, the multimodal Transformer would then be trained, validated, and tested. The PPG-BP dataset (219 patients) ~\cite{liang2018new}, consisting of PPG signals paired with their corresponding diabetes labels (37:182 in distribution), was also utilized for this experiment in order to evaluate the expansion and generalizability of the Performer model.

\subsection{PPG/ECG Data Processing and Training}

Since the above dataset came from various sources, they had different sampling frequencies. All medical waveforms were first resampled to a constant frequency of 128 Hz. The BioSPPY standard denoising filter was applied, along with a convolutional 1D layer to amplify the waveform patterns; the denoising and pattern amplification are critical to reduce the noise prevalent especially in the PPG signals. The data was then normalized to a range of [-1, 1] and was properly segmented to 4-second window frames with 2-second overlaps between each cut. Patches of different waveform lengths were also created to build the future hierarchical layers of SPA. 

\begin{table}
  \begin{center}
    {\small{
\begin{tabular}{l|l|l|l|l}
\midrule
CAD & \textbf{95.1\%} & 1.5\% & 1.8\% & 1.6\% \\
\midrule
CHF & 2.2\% & \textbf{96.5\%} & 0.5\% & 0.8\% \\
\midrule
MI & 1.8\% & 0.9\% & \textbf{96.2\%} & 1.1\% \\
\midrule
HoTN & 1.5\% & 1.6\% & 1.0\% & \textbf{95.9\%} \\
\midrule
& CAD & CHF & MI & HoTN \\
\bottomrule
\end{tabular}
}}
\end{center}
\caption{Confusion Matrix for Four CVDs on the MIMIC-III Dataset.}
\label{table2}
\end{table}

Performer was implemented via PyTorch, and the model was trained on an AWS Instance server with the configuration as AMI: Deep Learning; GPU: Tesla T4. We utilized Adam as the optimizer with the learning rate of 0.0001 and performed 50 epochs during the training.

\begin{table}
  \begin{center}
    {\small{
\begin{tabular}{l|l|l}
\midrule
Normal & \textbf{71.7\%} & 28.3\% \\
\midrule
Diabetes & 24.1\% & \textbf{75.9\%} \\
\midrule
& Normal & Diabetes \\
\bottomrule
\end{tabular}
}}
\end{center}
\caption{Confusion Matrix for Diabetes Detection.}
\label{table3}
\end{table}

\section{Performance and Medical Application}

In this section, we present our evaluation of the proposed Performer architecture with the SPA mechanism. We first use the root-mean-square error (RMSE) to measure the accuracy of PPG-to-ECG reconstruction and compare the results with other studies on the same dataset. Then, we create the confusion matrix to measure the CVD detection, evaluating the effectiveness of Performer as a solution for continuous cardiac monitoring. An ablation study is included to demonstrate the contribution of the multimodal Transformer through the utilization of our new digital biomarker. In addition, we evaluate the attention maps to present the effectiveness of SPA. Finally, a wearable prototype is proposed as a solution for continuous cardiac monitoring through Performer. 

\begin{table*}
  \begin{center}
    {\small{
\begin{tabular}{c|c c c c|c}
\toprule
& PPG Only & ECG Only & Reconstructed ECG Only & Original PPG \& ECG Pair & PPG \& Reconstructed ECG (Proposed) \\
\midrule
CAD & 35.6\% & 68.9\% & 79.5\% & 82.5\% & \textbf{95.1\%} \\
CHF & 36.8\% & 69.2\% & 81.6\% & 86.2\% & \textbf{96.5\%} \\
MI & 40.9\% & 72.5\% & 82.9\% & 88.1\% & \textbf{96.2\%} \\
HoTN & 38.6\% & 75.9\% & 80.1\% & 85.9\% & \textbf{95.9\%} \\
\bottomrule
\end{tabular}
}}
\end{center}
\caption{Ablation Study of Model Performance with Different Signal Inputs.}
\label{table4}
\end{table*}

\subsection{Evaluation of PPG-to-ECG Reconstruction}

We applied RMSE, calculated as follows, to measure the reconstruction results.

\[ RMSE = \sqrt{\frac{\sum_{i=1}^{N}{(Predicted_i-Actual_i)^2}}{N}} \]

In Table~\ref{table1} we compare the statistics of the RMSE scores between the traditional machine learning (ML) with feature-based models (DCT/XDJDL ~\cite{zhu2021learning}), the current deep learning (DL) algorithms (GAN ~\cite{sarkar2021cardiogan} and Encoder/Decoder ~\cite{chiu2020reconstructing}), the vanilla Transformer, and the proposed Performer model. The reconstruction Transformer achieved an RMSE of 0.29, surpassing the results of other related works largely due to its SPA mechanism.

\subsection{Evaluation of CVD Detection}

As shown in Table~\ref{table2}, the confusion matrix indicates the model performance in classifying and detecting the four CVDs on the MIMIC-III dataset. The Performer architecture achieved a high degree of accuracy (95\% confidence) by utilizing (1) self-attention to denoise waveforms, (2) multi-head and positional embeddings to incorporate both the sequences and relationships of the waveform into training, (3) SPA to optimize feature signal relationships, and (4) a multimodal framework to include both the PPG signals and its reconstructed ECG, as a new digital biomarker for cardiac disease prediction.

To measure the universal applicability of Performer, this study extended the performance analysis to include CVD-related diseases such as diabetes. As presented in Table~\ref{table3}, when our proposed architecture was applied to the PPG-BP dataset, the model generated diabetes detection results (95\% confidence) that were comparable with those of Avram \etal~\cite{avram2020digital}, whose team had a more cohesive dataset to base their model. Performer’s applicability to diabetes detection indicates that the model’s performance is consistent among different datasets, suggesting that Performer can provide a universal solution.

\subsection{Multimodal Transformer and the New Digital Biomarker}

In this section, we discuss the contribution of our multimodal Transformer and the new digital biomarker. To evaluate the benefits of the multimodal Transformer and the biomarker, we performed a basic ablation study to compare the CVD detection performance among the following signal inputs: raw PPG and ECG signals, reconstructed ECG only, pairs of raw PPG \& ECG, and the new digital biomarker generated by Performer. The results are shown in Table~\ref{table4}. Raw PPG signals showed the poorest performance, followed by ECG signals alone. It is interesting to observe that the reconstructed ECG signals yielded better results than the raw ECG signals. This was due to Performer’s embedment of PPG signal information into its reconstruction of the ECG, revealing additional signals and hidden patterns from the PPG/ECG reconstruction. The multimodal of raw PPG/ECG pairs produced the similar performance as the reconstructed ECG only signals. The PPG and reconstructed ECG achieved the most accurate results, indicating that using the novel digital biomarker through Performer yielded the best performance.

\subsection{Attention Map through SPA}

To visualize the model’s interpretation of the biomedical waveforms, Figure~\ref{figure5} shows the attention maps of the last layer of the architecture. The highlights around the QRS complex indicated that Performer effectively derived a “biological understanding” of the PPG and ECG waveforms through SPA, allowing its self-attention mechanism to directly focus on the key features. This ability is the combined result of (1) the multi-layer hierarchy design, (2) the effective computation and concatenation of sequences at different wavelengths, and (3) the shifted mechanism to successfully cover cross-patch connections.

\subsection{Medical Application: PEARL}

\begin{figure}[t]
\begin{center}
\includegraphics[width=0.9\columnwidth]{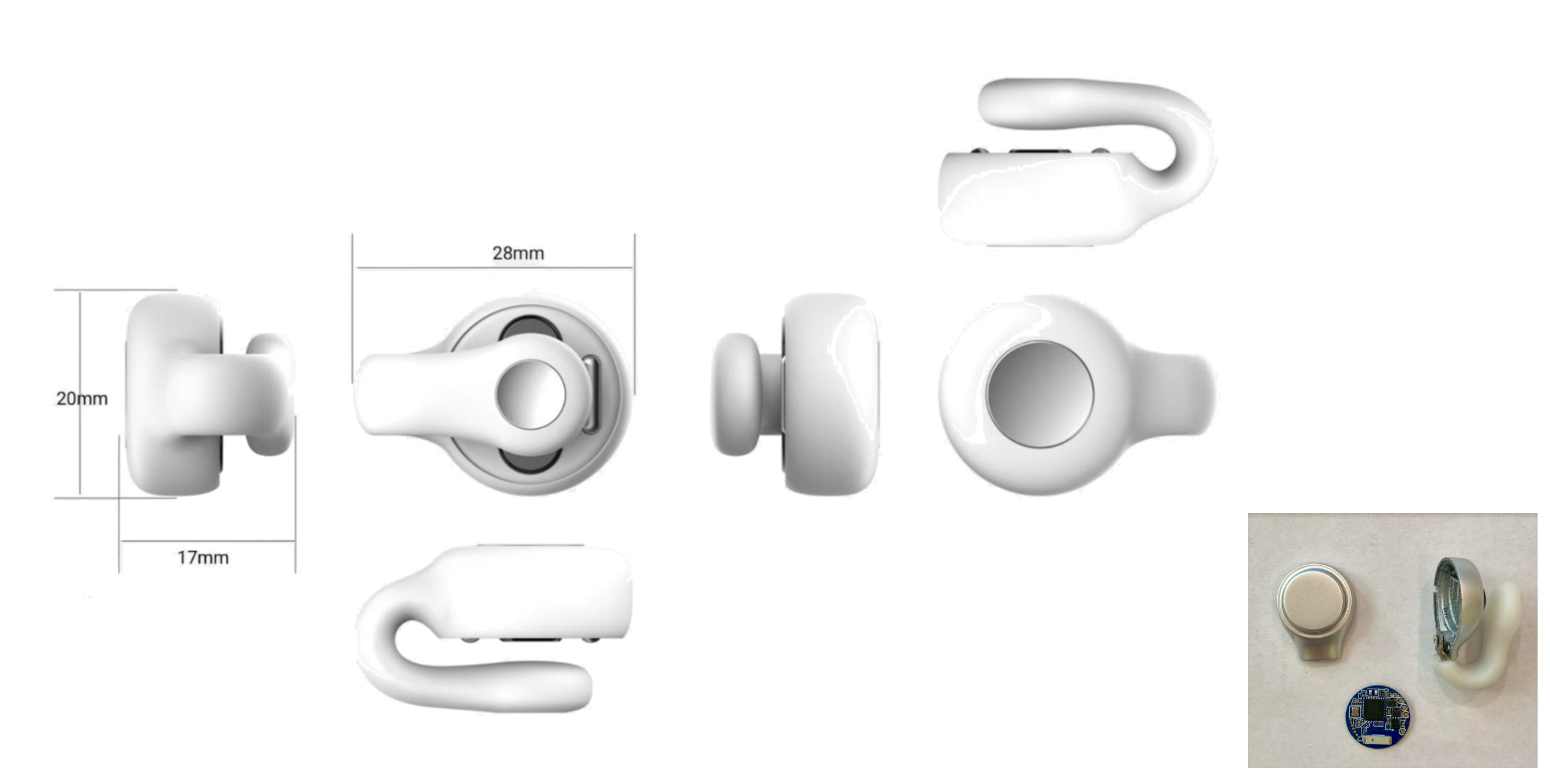}
\end{center}
   \caption{Exterior Design and Prototype of PEARL.}
\label{figure6}
\end{figure}

In addition to the architectural advancements presented in this paper, a healthcare wearable prototype was designed as a proof of concept for our proposed architecture. As seen in Figure~\ref{figure6}, PEARL – which stands for \textbf{P}PG \textbf{Ear}rings \textbf{L}ite – offers a potential solution for continuous cardiac monitoring without any active user participation. Positioning the wearable device on the earlobe is not only suitable for daily wearing but is also beneficial for producing accurate PPG waveform signals, given the minimal movement. PEARL sends the collected PPG signals to the Performer model in the Cloud, and the model returns any detected CVD risk back to the wearable. This application brings the real-world benefits of the proposed Performer architecture into life. Performer can also be used with other PPG-based wearables on the market, easily combining the value of ECG as a gold standard for cardiac monitoring through its PPG-to-ECG reconstruction.

\section{Conclusion and Future Work}

This research proposes Performer, a novel Transformer-based architecture that achieves state-of-the-art performance (0.29 RMSE) in PPG-to-ECG reconstruction, providing a solution for continuous cardiac monitoring by combining the advantages of both ECG/PPG signals. Furthermore, we create Shifted Patch-based Attention, an effective tokenization mechanism that capably serves as a general-purpose solution for biomedical waveform analysis via Transformers. Performer also demonstrated an encouraging performance for CVD detection by using PPG and reconstructed ECG signals. As a proof of concept, an earring wearable – PEARL – was designed on top of our proposed architecture.

Future work will explore the available longer range of PPG data through Performer. The range that spans from minutes and hours to days can reveal otherwise undiscovered patterns that would indicate CVD risk in the early phases, thereby reducing the casualties from CVD and CVD-related diseases. We also plan to experiment with reconstructing other biomedical signals via Performer, such as ballistocardiography (BCG) and phonocardiography (PCG), to broaden the usage of this architecture. Patients’ other medical records, such as their weight, age, blood pressure, and heart rate, can be incorporated into our future experimentations as well, to further improve Performer’s real-world applicability.

{\small
\bibliographystyle{ieee_fullname}
\bibliography{egbib}

\begin{thebibliography}{10}\itemsep=-1pt

\bibitem{avram2020digital}
Robert Avram, Jeffrey~E Olgin, Peter Kuhar, J~Weston Hughes, Gregory~M Marcus,
  Mark~J Pletcher, Kirstin Aschbacher, and Geoffrey~H Tison.
\newblock A digital biomarker of diabetes from smartphone-based vascular
  signals.
\newblock {\em Nature medicine}, 26(10):1576--1582, 2020.

\bibitem{bahdanau2014neural}
Dzmitry Bahdanau, Kyunghyun Cho, and Yoshua Bengio.
\newblock Neural machine translation by jointly learning to align and
  translate.
\newblock {\em arXiv preprint arXiv:1409.0473}, 2014.

\bibitem{banerjee2014photoecg}
Rohan Banerjee, Aniruddha Sinha, Anirban~Dutta Choudhury, and Aishwarya
  Visvanathan.
\newblock Photoecg: Photoplethysmographyto estimate ecg parameters.
\newblock In {\em 2014 IEEE International Conference on Acoustics, Speech and
  Signal Processing (ICASSP)}, pages 4404--4408. IEEE, 2014.

\bibitem{carion2020end}
Nicolas Carion, Francisco Massa, Gabriel Synnaeve, Nicolas Usunier, Alexander
  Kirillov, and Sergey Zagoruyko.
\newblock End-to-end object detection with transformers.
\newblock In {\em European conference on computer vision}, pages 213--229.
  Springer, 2020.

\bibitem{chiu2020reconstructing}
Hong-Yu Chiu, Hong-Han Shuai, and Paul C-P Chao.
\newblock Reconstructing qrs complex from ppg by transformed attentional neural
  networks.
\newblock {\em IEEE Sensors Journal}, 20(20):12374--12383, 2020.

\bibitem{dosovitskiy2020image}
Alexey Dosovitskiy, Lucas Beyer, Alexander Kolesnikov, Dirk Weissenborn,
  Xiaohua Zhai, Thomas Unterthiner, Mostafa Dehghani, Matthias Minderer, Georg
  Heigold, Sylvain Gelly, et~al.
\newblock An image is worth 16x16 words: Transformers for image recognition at
  scale.
\newblock {\em arXiv preprint arXiv:2010.11929}, 2020.

\bibitem{elgendi2019use}
Mohamed Elgendi, Richard Fletcher, Yongbo Liang, Newton Howard, Nigel~H Lovell,
  Derek Abbott, Kenneth Lim, and Rabab Ward.
\newblock The use of photoplethysmography for assessing hypertension.
\newblock {\em NPJ digital medicine}, 2(1):1--11, 2019.

\bibitem{gil2010photoplethysmography}
Eduardo Gil, Michele Orini, Raquel Bailon, Jos{\'e}~Mar{\'\i}a Vergara, Luca
  Mainardi, and Pablo Laguna.
\newblock Photoplethysmography pulse rate variability as a surrogate
  measurement of heart rate variability during non-stationary conditions.
\newblock {\em Physiological measurement}, 31(9):1271, 2010.

\bibitem{goldberger2000physiobank}
Ary~L Goldberger, Luis~AN Amaral, Leon Glass, Jeffrey~M Hausdorff, Plamen~Ch
  Ivanov, Roger~G Mark, Joseph~E Mietus, George~B Moody, Chung-Kang Peng, and
  H~Eugene Stanley.
\newblock Physiobank, physiotoolkit, and physionet: components of a new
  research resource for complex physiologic signals.
\newblock {\em circulation}, 101(23):e215--e220, 2000.

\bibitem{johnson2016mimic}
Alistair~EW Johnson, Tom~J Pollard, Lu Shen, Li-wei~H Lehman, Mengling Feng,
  Mohammad Ghassemi, Benjamin Moody, Peter Szolovits, Leo Anthony~Celi, and
  Roger~G Mark.
\newblock Mimic-iii, a freely accessible critical care database.
\newblock {\em Scientific data}, 3(1):1--9, 2016.

\bibitem{liang2018new}
Yongbo Liang, Zhencheng Chen, Guiyong Liu, and Mohamed Elgendi.
\newblock A new, short-recorded photoplethysmogram dataset for blood pressure
  monitoring in china.
\newblock {\em Scientific data}, 5(1):1--7, 2018.

\bibitem{lin2017feature}
Tsung-Yi Lin, Piotr Doll{\'a}r, Ross Girshick, Kaiming He, Bharath Hariharan,
  and Serge Belongie.
\newblock Feature pyramid networks for object detection.
\newblock In {\em Proceedings of the IEEE conference on computer vision and
  pattern recognition}, pages 2117--2125, 2017.

\bibitem{liu2012university}
David Liu, Matthias G{\"o}rges, and Simon~A Jenkins.
\newblock University of queensland vital signs dataset: development of an
  accessible repository of anesthesia patient monitoring data for research.
\newblock {\em Anesthesia \& Analgesia}, 114(3):584--589, 2012.

\bibitem{liu2021swin}
Ze Liu, Yutong Lin, Yue Cao, Han Hu, Yixuan Wei, Zheng Zhang, Stephen Lin, and
  Baining Guo.
\newblock Swin transformer: Hierarchical vision transformer using shifted
  windows.
\newblock In {\em Proceedings of the IEEE/CVF International Conference on
  Computer Vision}, pages 10012--10022, 2021.

\bibitem{reiss2019deep}
Attila Reiss, Ina Indlekofer, Philip Schmidt, and Kristof Van~Laerhoven.
\newblock Deep ppg: Large-scale heart rate estimation with convolutional neural
  networks.
\newblock {\em Sensors}, 19(14):3079, 2019.

\bibitem{sarkar2021cardiogan}
Pritam Sarkar and Ali Etemad.
\newblock Cardiogan: Attentive generative adversarial network with dual
  discriminators for synthesis of ecg from ppg.
\newblock In {\em Proceedings of the AAAI Conference on Artificial
  Intelligence}, volume~35, pages 488--496, 2021.

\bibitem{somani2021deep}
Sulaiman Somani, Adam~J Russak, Felix Richter, Shan Zhao, Akhil Vaid, Fayzan
  Chaudhry, Jessica~K De~Freitas, Nidhi Naik, Riccardo Miotto, Girish~N
  Nadkarni, et~al.
\newblock Deep learning and the electrocardiogram: review of the current
  state-of-the-art.
\newblock {\em EP Europace}, 23(8):1179--1191, 2021.

\bibitem{vaswani2017attention}
Ashish Vaswani, Noam Shazeer, Niki Parmar, Jakob Uszkoreit, Llion Jones,
  Aidan~N Gomez, {\L}ukasz Kaiser, and Illia Polosukhin.
\newblock Attention is all you need.
\newblock {\em Advances in neural information processing systems}, 30, 2017.

\bibitem{zhu2021learning}
Qiang Zhu, Xin Tian, Chau-Wai Wong, and Min Wu.
\newblock Learning your heart actions from pulse: Ecg waveform reconstruction
  from ppg.
\newblock {\em IEEE Internet of Things Journal}, 8(23):16734--16748, 2021.

\end{thebibliography}
}

\end{document}